
\magnification=\magstep1
\baselineskip=18pt
\overfullrule=0pt
\nopagenumbers
\footline={\ifnum\pageno>1\hfil\folio\hfil\else\hfil\fi}
\font\twelvebf=cmbx12

\rightline{CU-TP-613}
\rightline{August 1993}
\vskip 1in
\centerline{\twelvebf Self-intersection, axial anomaly and the string
             picture of QCD}
\vskip 1in
\centerline{\it Ravit Efraty}

\smallskip
\centerline{Physics Department, Columbia University}
\centerline{New York, NY 10027}
\vskip 1in
\centerline{\bf Abstract}
\smallskip
\noindent The leading, planar diagrams of the $1/N_c$ expansion
and the usual string description suggest that quarks propagate on the
boundary of a two-dimensional world surface. We restrict the quarks to the
boundary of the world surface by giving them infinitely large mass on the
interior of the surface and zero mass on its boundary and show that in this
picture the QCD $\theta$--vacua can be represented by the self-intersection
number (or equivalently by the first Chern number of the normal bundle)
of the surface.

\vskip 0.5in
\noindent This research was supported in part by the U.S. Department
of Energy
\vfill\eject

{\it 1.Introduction}
\vskip .2in
In recent years, an effort has been made to establish the relationship
between QCD and strings in order to better understand QCD. Indications
that strings might provide an effective description of QCD are seen in the
lattice formulation and the $1/N_c$ expansion [1,2].
In the limit, $N_c \rightarrow \infty ,$ only planar diagrams,
which have exactly the topology of the quantized string with quarks at its
ends, contribute. The two-dimensional structure obtained by attaching
surfaces to each loop in the planar diagram resembles the world sheet
 $\Sigma$ of an open string. If QCD behaves like a string theory,
we should find string theoretic representations of QCD characteristics.
One of the qualitatively important characteristics of QCD
is the existence of $\theta$--vacua. The string-candidate associated with the
$\theta$--term of QCD, ${\theta \over 16\pi^2} {\rm tr} F\tilde F,$ was
suggested in [3,4] to be the self-intersection number
of the surface, $I(\Sigma)$ [5].
In [4] Mazur and Nair introduced fermions
in the theory and projected the four-dimensional action
onto $\Sigma.$ This projection determines
the coupling of the fermions to the extrinsic and intrinsic
parameters of $\Sigma.$ They found that the effect of $\theta$--vacua,
represented in the QCD functional integral by $\exp (i \theta Q)$
where $Q$ is the instanton number, can be represented in the
string description by $\exp (i \theta C_1(\Sigma))$ where $C_1(\Sigma)$
is the first Chern number of the normal bundle to $\Sigma.$
However in the large $N_c$--limit description or the usual string description
the quarks are
constrained to the boundary of the two-dimensional surface, and in
[4] the quarks live on the entire surface. Therefore
a more accurate description of QCD would require a modified model.
In this work we restrict the quarks to the boundary by giving them
infinitely large mass on the interior of the surface and zero mass on the
boundary.
We calculate the axial anomaly for these
quarks with a coordinate-dependent mass
and obtain a similar relation between the first Chern number of the
surface and the total change of the axial charge, $\Delta Q^5.$

\vskip .2in
{\it 2.Short review}
\vskip .2in

Following Mazur and Nair [4], we begin with the four-dimensional
free fermion Lagrangian

$${\cal L} = \bar \Psi \gamma^a (\partial_a - {1\over 8} w_a^{\alpha\beta}
[\gamma_\alpha,\gamma_\beta])\Psi  \eqno(1)$$

\noindent where we have pulled back quantities from $\Sigma$
to $M$ and $w_a^{\alpha\beta}$ is the spin connection.
$M$ in general is a Riemann surface with real coordinates
$\xi_a \; (a= 1,2).$ The image of $M$ in ${\cal R}^4$ is the
surface $\Sigma.$ More than one point in $M$ can be mapped to
the same point in ${\cal R}^4$ corresponding to self-intersections
of $\Sigma.$ Thus the mapping $X^\mu(\xi): M \rightarrow {\cal R}^4 $
is, in general, only an immersion.
$w_a^{\alpha\beta}$ can be expressed in terms of the tangent and normal
vectors of the immersion $X^\mu(\xi)$ to give the Lagrangian for quarks
on $M.$
The tangent vectors are $t_a^\mu = \partial X^\mu / \partial \xi^a $
and the normal vectors are orthogonal to the tangents at the
image points on $\Sigma.$ The immersion of $M$ in ${\cal R}^4$
defines the tangent and normal bundles on $M.$ $A_{aAB}$ is the $SO(2)$
connection for the normal bundle. The extrinsic curvature $K^A_{ab}$ is
given by the equation

$$\partial_a\partial_b X^\mu = \Gamma^c_{ab}\partial_c X^\mu
+K^A_{ab} n_A $$

\noindent where $\Gamma^c_{ab}$ is the usual Christoffel symbol for the induced
metric, $g_{ab} = \partial X^\mu / \partial \xi^a \; \partial X^\mu / \partial
 \xi^b .$
The frame fields on $M, \; e_a^{\hat a},$ are given by
$e_a^{\hat a}e_{\hat a b}=g_{ab}.$ The target space is
 ${\cal R}^4,$ thus the zero-curvature spin connection is

$$w_a^{\alpha\beta} =t_a^\nu E^{\alpha\mu}\nabla_\nu E^{\beta\mu} $$

\noindent where $ E^{\alpha\mu}$ decomposes as follows

$$ E^{\alpha\mu} = (n^{A\mu}, t^\mu_{\hat a}e^{a \hat a}).$$

\noindent Hence the Lagrangian for fermions on $M$ may be written as

$${\cal L} = \bar \Psi \gamma^a (\partial_a - {1\over 8}\Gamma^c_{ab}
[\gamma^b,\gamma_c] +{1 \over 8} A_{aAB} [\gamma^A,\gamma^B] -
{1\over 4} K^A_{ab} [\gamma_b,\gamma_A])\Psi.  $$

\noindent Note that this Lagrangian only contains geometric information.
The strings description of QCD should incorporate confinement,
thus to obtain a purely geometrical coupling for color singlet
combinations there should be an extra factor of $1/N_c$ in the couplings
of $K^A_{ab}$ and $A_{aAB}$ to quarks. This factor arises naturally in the
large $N_c$--limit of QCD since the quark contribution is smaller by a factor
of $1/N_c$ relative to the gluon contribution.
Thus the Lagrangian for quarks on $M$ can be taken to be

$${\cal L} = \bar q^i \gamma^a D_a q_i$$

\noindent where $D_a$ is given by

$$ D_a= \partial_a - {1\over 8}\Gamma^c_{ab}
[\gamma^b,\gamma_c] +{1 \over 8 N_c} A_{aAB} [\gamma^A,\gamma^B] -
{1\over 4 N_c} K^A_{ab} [\gamma_b,\gamma_A] . \eqno(2) $$

\noindent The standard heat kernel method can be used to obtain the axial
anomaly associated with the covariant derivative $D_a.$ The functional trace
of $\gamma_5,$ the quantity equal to the instanton number in four-dimensional
QCD, was found to be [4]

$${\rm Tr} \gamma_5 = {1 \over  N_c} C_1(\Sigma) \eqno(3) $$

\noindent where $C_1(\Sigma),$ the first Chern number, is related
to the self intersection number $I(\Sigma)$ by

$${1 \over 2} C_1(\Sigma) = I(\Sigma) \equiv - {1\over 16 \pi} \int d^2\xi
\sqrt{g}\,
 g^ab \nabla_a t^{\mu\nu}\nabla_b \tilde t^{\mu\nu},$$

\noindent and has the simple form

$$ C_1(\Sigma) = {1 \over 2\pi} \int_M {\rm tr} F. \eqno(4)$$

\noindent $F$ is the field strength tensor for the $SO(2)$ normal
bundle connection and $t^{\mu\nu}= {\epsilon^{ab}\over \sqrt{g}}t^\mu_a
t^\nu_b.$

A complete analogy with QCD is lacking here because in this picture
the quarks live on the two-dimensional surface, rather then
propagating only on the boundary as is required by the leading
order diagrams of the $1/N_c$ expansion and by the usual string
description.
To provide a better analogy we instead choose a regularization
scheme which allows us to compute the anomaly in two dimensions,
while restricting the quarks to the boundary of the surface.

\vskip .7in
{\it 3.Regularization scheme}
\vskip .2in

We give the quarks  zero mass on the boundary and infinitely large
mass inside the surface. We shall Calculate the propagator for these
coordinate-dependent mass quarks and prove that they are indeed
restricted to the boundary.
For simplicity we take space to be flat with the spatial direction
$\sigma$ varying between $-l$ and $l.$
$m(\sigma),$ the quark mass, is chosen to have the simple form:

$$m(\sigma)= \cases{0,& $-l\leq \sigma \leq -a \;\;\; {\rm region\; I}$;\cr
               m_0,& $-a< \sigma < a \;\;\; {\rm region\; II}$ ;\cr
               0,& $a\leq \sigma \leq l \;\;\; {\rm region\; III}$ .\cr}
               \eqno(5)$$

\noindent We are interested in the limit $m_0 \rightarrow \infty.$
We shall assume that the gauge field of interest, viz. $(F_{ab})_{AB},$
the field-strength for $A_{aAB},$ is slowly varying on the scale of
$(l-a).$

\vskip .2in
{\it 4.Calculating the anomaly for a coordinate-dependent mass }
\vskip .2in

In order to derive an equation similar to equation (3)
we calculate the quantity $\Delta Q^5$ given by

$$ \Delta Q^5 = \int \partial_\mu J^{5\mu} d^2\xi \eqno(6)$$

\noindent where $J^{5\mu} $ is the neutral gauge invariant axial current,
$J^{5\mu}  = {\bar \Psi} \gamma^\mu \gamma^5 \Psi, $ and

$$\partial_\mu J^{5\mu} = {\cal A} + 2 i m(x) {\bar \Psi(x)} \gamma^5
\Psi(x) \eqno(7)$$

\noindent where ${\cal A}$ is the anomalous term. There are different ways to
calculate the axial anomaly.
The perturbative method is best suited to our choice of mass.
The two-dimensional axial anomaly stems from the two-point
function

$$ T^5_{\mu\nu}  = \langle T(J_\mu(y) J^5_\nu(x)) \rangle. $$

\noindent For a particle with constant mass, $m,$ requiring $J_\mu^5 (x)$
to be gauge invariant results in an anomalous term, $\rho_2$:

$$\rho_2 = -{e \over 2\pi} \epsilon^{\mu\nu}F_{\mu\nu}(x). \eqno(8)$$

\noindent Since we want to calculate $q^\nu T_{\mu\nu}^5 $ in a spatially
confined space we choose a cutoff regularization.
We are also interested in the limit $m_0 \rightarrow \infty,$ thus
we choose the cutoff $\Lambda$ and the fixed mass $m_0$ such
that

$$ \Lambda < m_0 . \eqno(9)$$

\noindent We next compute the propagator for $m = m(\sigma).$
In view of equation (9)  we search for the solution to the Dirac
equation:

$$ (\gamma \cdot \partial - m(\sigma) ) \Psi (\sigma,t) = 0$$

\noindent with $|E| < \Lambda < m_0$ which satisfy the boundary conditions
required for a self-adjoint hamiltonian, viz:

$$\eqalign{\Psi_{\rm I}^*(-l)\gamma^0\gamma^1\Psi_{\rm I}(-l)&=0 \cr
           \Psi_{\rm III}^*(l)\gamma^0\gamma^1\Psi_{\rm III}(l)&=0\cr}
           \eqno(10)$$

\noindent and the matching conditions at $\sigma = \pm a,$
where $\Psi_i $ is the solution in region $i.$
Separation of the time-variable requires a two-phase ansatz:

$$\Psi = e^{-iEt}\Psi^{(+)}(\sigma) + e^{iEt}\Psi^{(-)}(\sigma).$$

\noindent The static functions $\Psi^{\pm}(\sigma)$ satisfy

$$ \pm E \Psi^{(\pm)}(\sigma)=(\alpha p + \beta m)\Psi^{(\pm)}(\sigma).$$

\noindent Our representation for the Dirac matrices is
$ \alpha = \sigma^2,  \, \beta = \sigma^1.$
Implementing the matching conditions and requiring
a normalizable solution when $m_0 \rightarrow \infty,$ while
fixing the wave function in region III  (or I) gives a solution
of exponentially decaying wave functions in regions I (or III) and II,
but an oscillating wave function in region III (or I).
The zero flux condition at $x=l$ gives quantization of energy

$$ 2k_n (l-a)+ \varphi = 2 \pi n \;\;\; \{ n \in {\cal Z} : |k_n| <
                           \Lambda \} \eqno(11)$$

\noindent where $k^2 = E^2 $ and
$\sin \varphi = {k \over m_0}.$

\noindent For large $m_0,$ $\sin\varphi \approx \varphi$ gives

$$k_n = {\pi n \over (l-a) + {1 \over 2m_0} }
\;\;\; -N \leq n \leq N \eqno(12)$$
\noindent where
$$  |N| \leq {\Lambda \over \pi} ({1 \over 2m_0} + (l-a)).$$

\noindent Fixing $\Psi_{\rm I}$ instead of $\Psi_{\rm III}$ yields

$$ 2k_n (l-a)+ \varphi = 2 \pi (n+1/2) \;\;\;  \{n \in { \cal Z} :\, |k_n|
< \Lambda\}. \eqno(13)$$

\noindent The other boundary condition is automatically fulfilled in the large
$m_0$ limit.
The propagator is constructed from $\Psi^{(\pm)}(x)$

$$\eqalign{S(x, x^\prime) =& -i \theta (t^\prime - t) \sum_{n=-N}^N
e^{-iE_n(t^\prime-t)}
\Psi_n^{(+)}(x^\prime){\bar \Psi_n^{(+)}(x)}\cr
&+ i \theta(t-t^\prime) \sum_{n=-N}^N
e^{iE_n(t^\prime-t)} \Psi_n^{(-)}(x^\prime){\bar \Psi_n^{(-)}(x)}.\cr}
\eqno(14)$$

\noindent For $k_n$ satisfying (12) the propagator in regions I
and II decays exponentially. Note that in region III the propagator
describes a Dirac particle in an infinite square potential well
of width $(l-a)+{1 \over 2m_0}.$

The propagator of a massless fermion in an infinite square potential
well of width $(l-a),$ denoted by $S^0(x^\prime,x),$ is given by
equation (14) with

$$ \Psi_n^{(+)}(x^\prime){\bar \Psi_n^{(+)}(x)} = -{1\over l-a}
\pmatrix{{k_n \over E_n}\sin k_n(x^\prime-l)\cos k_n(x-l)&
\sin k_n(x^\prime-l)\sin k_n(x-l)\cr
         \cos k_n(x^\prime-l)\cos k_n(x-l)&
{k_n \over E_n}\cos k_n(x^\prime-l)\sin k_n(x-l)\cr}
         \eqno(15)$$

\noindent where

$$k_n = {\pi n \over l-a} \;\;\;\; N^0 \leq {\Lambda \over \pi}(l-a)
 \;\;\;\; -N^0 \leq n \leq N^0.  \eqno(16)$$

\noindent $\Psi_n^{(-)}(x^\prime){\bar \Psi_n^{(-)}(x)}$ is obtained from
(15) by replacing $E_n$ with $-E_n.$
\noindent $S^0(x^\prime,x)$ does not have the usual $x^\prime -x$ dependence.
This is not surprising since in confining the quarks we have
lost translational invariance. It is
straightforward to check that $S^0(x^\prime,x)$ is indeed the propagator.
The propagator corresponding to a well of width $(l-a)+{1 \over 2m_0}$
differs from $S^0(x^\prime,x)$ by terms of order ${1 \over m_0}$
which appear in the normalization of the wave function, $\Psi_{\rm III},$

$$ |A|^2 = {1 \over 4(l-a)}(1 - {1 \over (l-a) 2m_0})$$

\noindent and in the argument of the trigonometric functions

$$k_n = {\pi n \over 2 (l-a)}(1 - {1 \over (l-a) 2m_0}).$$

\noindent (Though the interval in equation (12) appears different than that
in equation (16), we are free to choose $N = N^0.$)
When computing $T_{\mu\nu}^5$ in region III, no additional factors of $m_0$
appear. This insures that the $m_0 \rightarrow \infty$ limit gives the same
expression for $T_{\mu\nu}^5$ as in the massless case. Thus we find for
region III:

$${\cal A} = \rho_2. \eqno(17)$$

\noindent A similar argument for $k_n = {\pi (n+1/2) \over
(l-a) + 1/2m_0}$ gives the usual two-dimensional anomaly in
region I.
In region II, using the cutoff regularization, the propagator  between
$\sigma$ and $\sigma^\prime$ decays exponentially. Therefore all diagrams in
that region vanish in the large $m_0$ limit. Also all matrix elements
of $J^5 = m_0 {\bar \Psi}
\gamma^5\Psi$ vanish in this regularization scheme since the factor of $m_0$
cannot compensate for the exponential decay. This gives in region II,

$$\partial_\mu J^{5\mu} = 0  , \eqno(18)$$

\noindent which is expected. A simple way to see this is to recall that
for a massive fermion in two dimensions
we have

 $$\partial_\mu J^{5\mu}(x) = \rho_2 + 2i m{\bar \Psi}(x) \gamma^5\Psi(x).$$

\noindent When $m \rightarrow \infty$ the matrix elements of ${\bar \Psi}
\gamma^5\Psi$ behave like ${1\over m},$ resulting in a non-vanishing
remainder for $2i m{\bar \Psi} \gamma^5\Psi.$ In fact this
remainder is exactly $-\rho_2.$ Thus we get for a heavy fermion

$$\lim_{m\rightarrow \infty} \partial_\mu J^{5\mu}(x)  = 0$$

\noindent which is the same result as (18).

\noindent (To understand this better, recall [6] that another way to assure
gauge invariance is by introducing a Pauli-Villars regulator field $\psi$
with mass $M$ and a regulated axial current. Then
$\partial_\mu J^{5\mu}(x) = 2i m{\bar \Psi}(x) \gamma^5\Psi(x)
-2i M{\bar \psi}(x) \gamma^5\psi(x).$
When $M \rightarrow \infty$ the regulator field contribution gives
the axial anomaly.)

To summarize

$$\partial_\mu J^{5\mu} = \cases{\rho_2,& $-l\leq \sigma \leq -a$;\cr
               0,& $-a< \sigma < a$;\cr
               \rho_2,& $a\leq \sigma \leq l$.\cr} \eqno(19)$$

\vskip .2in
{\it 5.Conclusion}
\vskip .2in
The total change of the axial charge, $\Delta Q^5,$
constructed from the axial current is

$$ \Delta Q^5 = {1 \over N_c}\int dt \int_{\rm I+III}d\sigma
\rho_2 \eqno(20)$$

\noindent where the same argument that led to the covariant derivative
of equation (3) results in a coefficient of $1/N_c.$
It is this quantity that we want to relate to the
first Chern number or to the self-intersection
number of the surface $\Sigma.$ (In [4] a relation to the functional
trace of $\gamma^5$ emerges naturally from the functional integral
formalism.)
In equation (4) $C_1(\Sigma)$ is expressed as
an integral of the field strength tensor, $ F_{abAB}. $
Since the self-intersection number is a topologically invariant quantity
of the surface $\Sigma,$ given a connection configuration $A_{aAB},$
we can smoothly deform it to $A^{\prime}_{aAB},$ such that

$${1\over 8 \pi} \int \epsilon^{AB}
\epsilon^{ab} (F_{ab})_{AB} d^2\xi  = {1\over 8 \pi} \int \epsilon^{AB}
\epsilon^{ab} (F^{\prime}_{ab})_{AB} d^2\xi.  $$

\noindent We can choose this deformation to be the one for which the field
strength
tensor $ F_{abAB}$ vanishes in region II, i.e. where
$m(\sigma) \neq 0. $  This insures that the
normal bundle is trivial in region II hence only regions I and III contribute
to the self-intersection number. Thus we can write

$$C_1(\Sigma) = 2I(\Sigma) = {1\over 8 \pi} \int_{\rm I+III} \epsilon^{AB}
\epsilon^{ab} (F_{ab})_{AB} d^2\xi. \eqno(21) $$

\noindent The integrand is exactly $\rho_2.$ In both equations
(20) and (21), the area of integration, which is the `dynamic' area,
is limited to a strip along the boundary where the quarks live.
This allows us to relate the QCD topological charge to the first Chern number
in the string description

$$ {1 \over N_c} C_1 (\Sigma) = \Delta Q^5, \eqno(22) $$

\noindent in the same manner as in ref. [4], but for a more accurate
picture in which the quarks are restricted to the boundary.

\noindent Note that if we deform $A_{aAB}$ such that a self-intersection
occurs in region II, the gauge fields are no longer varying slowly on the
scale of $(l-a).$ This must be taken into account when deriving the
anomaly (e.g. requiring a more careful treatment of the boundaries
between the regions). Presumably, a non-vanishing
divergence of the axial current in region II will result
and equation (22) will still be valid.

\vskip .2in
{\it Acknowledgements}
\vskip .2in
It is a pleasure to thank V.~P.~Nair for discussions at every stage of this
work and a critical reading of the manuscript.

\vskip .2in
{\bf References}
\vskip .2in
\item{[1]} G.~'t Hooft, {\sl Nucl. Phys.} {\bf B72} (1974) 461;
       {\bf B75} (1974)
      461.
\item{[2]} V.~A.~Kazakov, {\sl Sov. Phys. JETP} {\bf 58(6)} (1983) 1096.
\item{[3]} A.~P.~Balachandran, F.~Lizzi and G.~Sparano, {\sl Nucl. Phys.}
      {\bf B263} (1986) 608. A.~Polyakov, {\sl Nucl. Phys.}
      {\bf B268} (1986) 406.
\item{[4]} P.~O.~Mazur and V.~P.~Nair, {\sl Nucl. Phys.}
      {\bf B284} (1986) 146.
\item{[5]} For general background see M.~Spivak,
    {\sl A Comprehensive introduction to differential geometry},
    (Publish of Perish, Inc, Berkeley, 1979).
\item{[6]} R.~Jackiw in {\sl Current algebra and anomalies} by
     S.~B.~Treiman, R.~Jackiw, B.~Zumino and E.~Witten, (Princeton
     University Press, Princeton, New Jersey, 1985).
\bye